\documentclass{article}
\usepackage{ijcai20}

\usepackage{todonotes}

\usepackage{times}

\usepackage{soul}
\usepackage{url}
\usepackage[hidelinks]{hyperref}
\usepackage[utf8]{inputenc}
\usepackage[small]{caption}
\usepackage{subcaption}
\usepackage{graphicx}
\usepackage{amsmath}
\usepackage{booktabs}
\usepackage{dblfloatfix}
\usepackage{todonotes}
\title{Using an interpretable Machine Learning approach to study the drivers of International Migration}

\author{
Harold Silvère Kiossou$^1$\footnote{Contact Author}\and
Yannik Schenk $^1$\and
Frédéric Docquier $^{1}$ \and
Vinasetan Ratheil Houndji$^{1,2}$\and
Siegfried Nijssen$^1$\and
Pierre Schaus$^1$ \\
\affiliations
$^1$UCLouvain, Belgium\\
$^2$UAC, Benin\\
\emails
\{harold.kiossou, yannik.schenk, frederic.docquier, vinasetan.houndji, siegfried.nijssen, pierre.schaus\}@uclouvain.be}

\begin{document}
\maketitle
\begin{abstract}
Globally increasing migration pressures call for new modelling approaches in order to design effective policies. It is important to have not only efficient models to predict migration flows but also to understand how specific parameters influence these flows.
In this paper, we propose an artificial neural network (ANN) to model international migration.
Moreover, we use a technique for interpreting machine learning models, namely Partial Dependence Plots (PDP), to show that one can well study the effects of drivers behind international migration. 
We train and evaluate the model on a dataset containing annual international bilateral migration from $1960$ to $2010$ from $175$ origin countries to $33$ mainly OECD destinations, along with the main determinants as identified in the migration literature.
The experiments carried out confirm that: 1) the ANN model is more efficient w.r.t. a traditional model, and 2) using PDP we are able to gain additional insights on the specific effects of the migration drivers. 
This approach provides much more information than only using the feature importance information used in previous works.
\end{abstract}

\section{Introduction}
\label{sec:intro}
Today  more people are on the move than ever before. The current number of international migrants, defined as individuals residing outside of their country of birth, is estimated to be almost 272 million people globally, representing 3.3\% of the world's population \cite{desa2019world}.
Drivers of migration, voluntary or forced, are numerous and often interrelated. Factors linked to economic prosperity, inequality, demography, conflict and  persecution, environmental change and natural disasters create a complex interplay of migration incentives and opportunities. On the other hand, migration flows have large implications for receiving countries, impacting economic and demographic structure, culture, environment, spread of infectious diseases, etc. Against this context, deepening the understanding of \textit{which} factors determine international migration and \textit{how} they do so is crucial to build effective governance strategies and policies.

Modeling human mobility is usually done using extended gravity models \cite{zipfP1P2Hypothesis1946,letouzRevisitingMigrationDevelopmentNexus2009,andersonGravityModel2011} or radiation models \cite{siminiUniversalModelMobility2012,masucciGravityRadiationModels2013}.  Both are analytical methods which rely on linear or log-linear relationships between the independents (migration flows) and a set of covariates (features). 
While these methods often attempt to identify causal effects, they lack predictive power and struggle to match important non-linearities in the migration nexus.


In this paper, we combine a long series of \textit{annual} flows of international migrants with a full set of the key migration drivers identified in previous research. Our contribution is twofold. First, we use this unique data set to show that our model of artificial neural networks (ANN) performs well in estimating yearly migration flows and outperforms the canonical gravity model. 
Second, we demonstrate how partial dependence plots as an ML interpretability technique can be used to explore complex non-linearities \cite{friedmanGreedyFunctionApproximation2001,molnarInterpretableMachineLearning2018}, allowing for insights that go beyond traditional radiation or gravity models. 
While we do not claim causality, our results highlight the potential of ML-based techniques as an exploratory tool to enrich the usual econometric analysis in a variety of fields.

The remainder of this paper is organized as follows. We first describe some related work as well as the data used in Section~\ref{sec:background}. 
Then, in Section~\ref{sec:migration_model}, we explain our model and the learning process. 
Finally, we show the results of the experiments carried out in Section~\ref{sec:experiments}.

\section{Background}
\label{sec:background}
\begin{table*}
    \centering
    \begin{tabular}{@{}lll@{}}
        \toprule
        \textbf{Input features$_{i, j, t}$} & & \textbf{Description} \\ 
        \midrule
        $gdpc_{i,t}$ & & Gross Domestic Product per Capita of origin country $i$ during the year $t$ \\
        $gdpc_{j,t}$ & & Gross Domestic Product per Capita of destination country $j$ during the year $t$ \\
        $pop_{i,t}$ & & Population size of origin country $i$ during year $t$\\
        $pop_{j,t}$ & & Population size of destination country $j$ during year $t$ \\
        $density_{i,t}$ & &  Population per $km^2$ at origin country $i$ during year $t$ \\
        $density_{j,t}$ & &  Population per $km^2$ at destination country $j$ during year $t$ \\
        $dep\_ratio_{i,t}$ & &  Dependency ratio at origin country $i$ during year $t$ \\
        $dep\_ratio_{j,t}$ & &  Dependency ratio at destination country $j$ during year $t$ \\
        $inttot_{i,t}$ && Magnitude score of episode of international warfare
involving origin country $i$ during year $t$ \\
        $civtot_{i,t}$ && Magnitude score of episode of civil warfare involving origin country $i$ during year $t$ \\
        $disaster_{i,t}$ && Natural disaster event in origin country $i$ during year $t$ \\
        $disaster_{j,t}$ && Natural disaster event in origin country $j$ during year $t$ \\
        $Stock_{i,j,d}$ && Migrant stock from $i$ living in $j$ at the beginning of decade $d$ \\
        $distw_{i,j}$ && Population weighted distance between origin country $i$ and destination country $j$ \\
        $comlang\_off_{i,j,t}$ & & Common official language between origin country $i$ and destination country $j$ during year $t$ \\
        $comrelig_{i,j,t}$ && Share of people with the same religion between origin country $i$ and destination country $j$  during year $t$ \\
        $droughts_{i,t}$ && Drought event at origin country $i$ during year $t$ \\
        $droughts_{j,t}$ && Drought event at destination country $j$ during year $t$ \\
        $T_{i,j,t}$ & & Migration flow from country $i$ to country $j$ during year $t$ \\
        \bottomrule
    \end{tabular}
        \caption{The input features used for the model. Each feature spans from 1960 to 2010 for a pair of origin-destination country.}
    
    \label{tab:features}
\end{table*}

The aim of predicting human mobility is to estimate the migrant flow $T_{i,j}$ from an origin country $i$ to a destination country $j$. Let $\hat{T}_{i,j}$ denote this estimation.

\paragraph{Related works.} 
Estimating migration flows using ML tools is a recent approach. To our knowledge, the only work published on this is \citeauthor{robinsonMachineLearningApproach2017}~\shortcite{robinsonMachineLearningApproach2017}.

The authors implement two ML-based techniques to forecast internal and international migration: the \emph{``extreme'' gradient boosting regression (XGBoost)} model, and a deep learning based \emph{artificial neural network (ANN)} model to estimate $\hat{T}_{i,j} = f(features)$ from a set of features. They estimate internal migration across US counties and, closer related to our analysis, changes in bilateral international migrant stocks over decades. In addition, based on the trained XGboost model, they  identify the most important features affecting migration, ordering them by their percentage of importance. By doing so, one can identify the drivers that mostly influence the model's predictions. 
Unfortunately, this method does not allow to understand \textit{how} these features influence migration flows.
Further, the low frequency data on international migration used makes it difficult to derive insights that are policy relevant in the short-run. Estimates may fail to account for the timing of events causing \textit{direct} migratory responses, such as natural disasters or conflicts, and miss a potentially important share of temporary migrants.


\paragraph{Our approach.} 
As in \citeauthor{robinsonMachineLearningApproach2017}~\shortcite{robinsonMachineLearningApproach2017}, we want to predict $\hat{T}_{i,j}$ from a set of features.
However, here, we propose a ML model that is able to estimate international human mobility on an annual basis.

To capture the complexity behind international migration dynamics, we add a set of different features of sending and receiving countries (conflict, disaster, etc.) and importantly account for the existence of initial migrant networks.

We analyze the covariates (features) in detail with interpretable ML techniques that go beyond the typical ``black box'' estimates. Our approach is based on partial dependence plots, allowing us to gain additional insights regarding international migration patterns, especially in the presence of strongly non-linear effects.

\begin{table*}
\centering
\begin{tabular}{@{}lrrrrrrrrrrrr@{}}
    \toprule
    & \multicolumn{2}{c}{CPC} & & \multicolumn{2}{c}{MAE} & & \multicolumn{2}{c}{RMSE} & & \multicolumn{2}{c}{$r^2$} \\\cmidrule{2-3} \cmidrule{5-6} \cmidrule{8-9} \cmidrule{11-12}
    Models & train & test & & train & test & & train & test & & train & test \\\midrule
    PPML & 0.537 & 0.591 & & 1157 & 819 & & 8211 & 4230 & & 0.237 & 0.481\\
    ANN & \textbf{0.776}  & \textbf{0.617} & & \textbf{545} & \textbf{724} & & \textbf{5434} & \textbf{4088} & & \textbf{0.670} & \textbf{0.520}\\
    \bottomrule
\end{tabular}
\caption{Comparison of the 2 models for the specified metrics. The values shown are by pair (train - test).}
\label{tab:metrics_cmp}
\end{table*}

\paragraph{Data and features set.}
Table ~\ref{tab:features} describes the features used to train and test our model. The dataset used is a combination of data from several sources.
We combine data from the Determinants of International Migration C2C database (DEMIG) of historic migration flows \cite{vezzoli2014uncovering}, with the more recent International Migration Data database (IMD) \cite{stat2009international}. To the best of our knowledge, this leads to the most comprehensive data set on annual bilateral migration flows taken to analyse in this context, covering flows from $175$ origins to $33$ mainly OECD destinations over the years $1960$ to $2010$\footnote{ We limit the analysis to the period 1960-2010 to avoid the turmoil related to the recent refugee crisis that took of with the beginning of the Syrian civil war in 2011.}. Both data sets are based on population registers and residence permits, and are strongly coherent with a Pearson correlation coefficient of over 0.99 for the observations reported in both sources. We draw on the IMD data as primary source and append missing observations with DEMIG data.

The features used as independent variables in the model can be split in two main categories: 
(a) country specific indicators such as population and population density, GDP per capita, dependency ratio\footnote{Dependency ratios are defined as the ratio between young ($<15$) and old ($>65$) to the working age population.}, internal and international conflicts, natural disasters (earthquakes, storms, floods, volcanic eruptions) and droughts\footnote{We use gridded temperature and rainfall data to calculate SPEI-based relevant yearly drought indicators by country \cite{begueria2014standardized}} for both origin and destination countries at each time step; 
and (b) bilateral indicators containing distances between origin and destination at time, e.g.  geographical distance, common language and common religion.
Finally, migrant networks have been shown to be important drivers of subsequent migration flows \cite{beine2009diasporas}. We account for the presence of initial migrant networks by adding the stock of migrants born in $i$ and living in $j$ at the beginning of each decade to our feature set. The corresponding data stems from the Global Bilateral Migration Database (GBMD) \cite{ozdenWhereEarthEverybody2011}.




\section{Our ML model}
\label{sec:migration_model}
\begin{figure*}
    \centering
        \begin{subfigure}[b]{0.305\textwidth} 
        \centering \includegraphics[width=\textwidth]{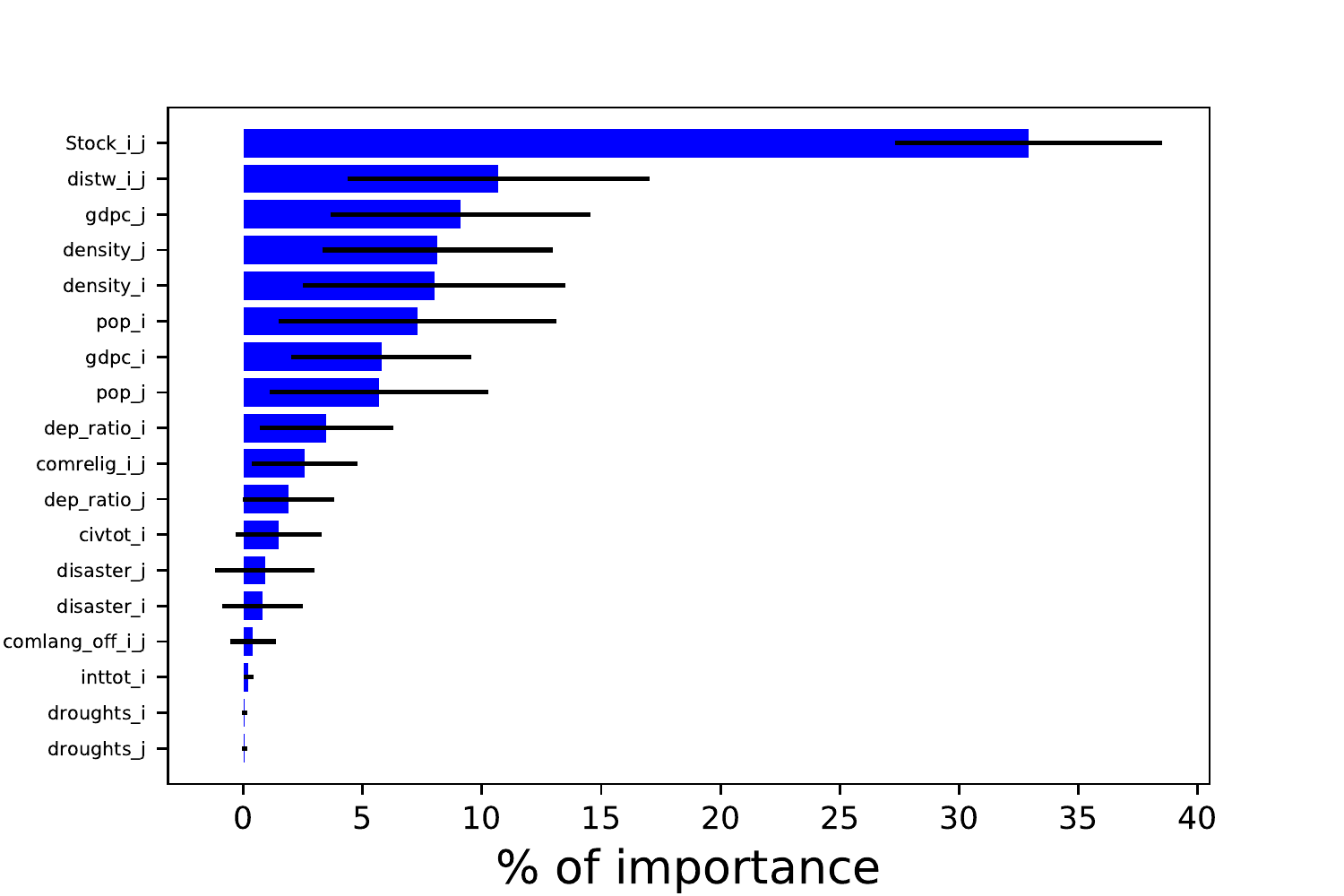}
        \caption{Features importance using RF.}
        \label{fig:features_imp}
    \end{subfigure}
    ~
    \begin{subfigure}[b]{0.295\textwidth} 
        \centering \includegraphics[width=\textwidth]{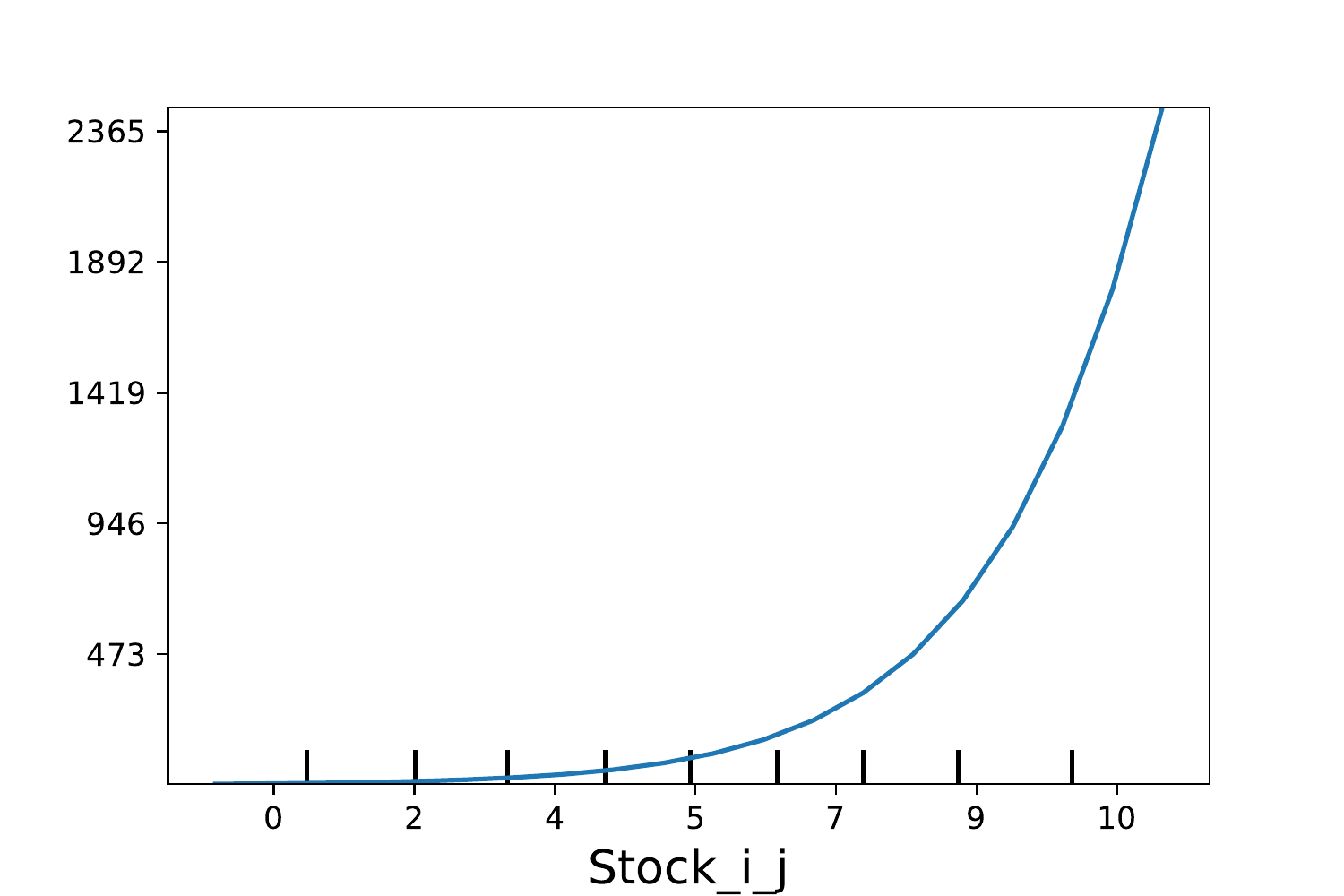}
        \caption{PDP for $Stock_{i,j,t}$.}
        \label{fig:pdp_stock}
    \end{subfigure}
    ~ 
    \begin{subfigure}[b]{0.30\textwidth}
        \centering \includegraphics[width=\textwidth]{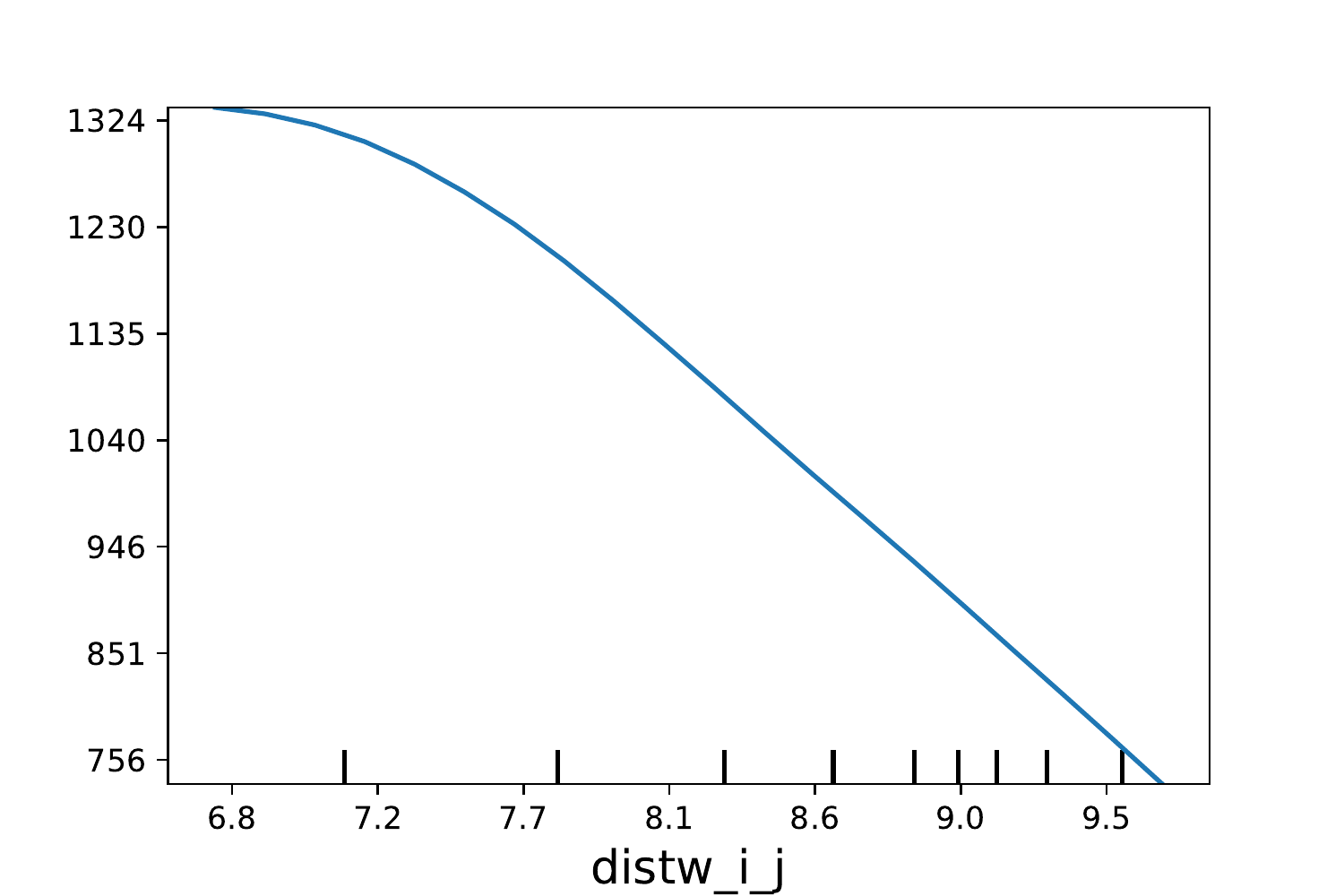}
        \caption{PDP for $distw_{i,j,t}$.}
        \label{fig:pdp_distw}
    \end{subfigure}

    \begin{subfigure}[b]{0.30\textwidth}
        \centering \includegraphics[width=\textwidth]{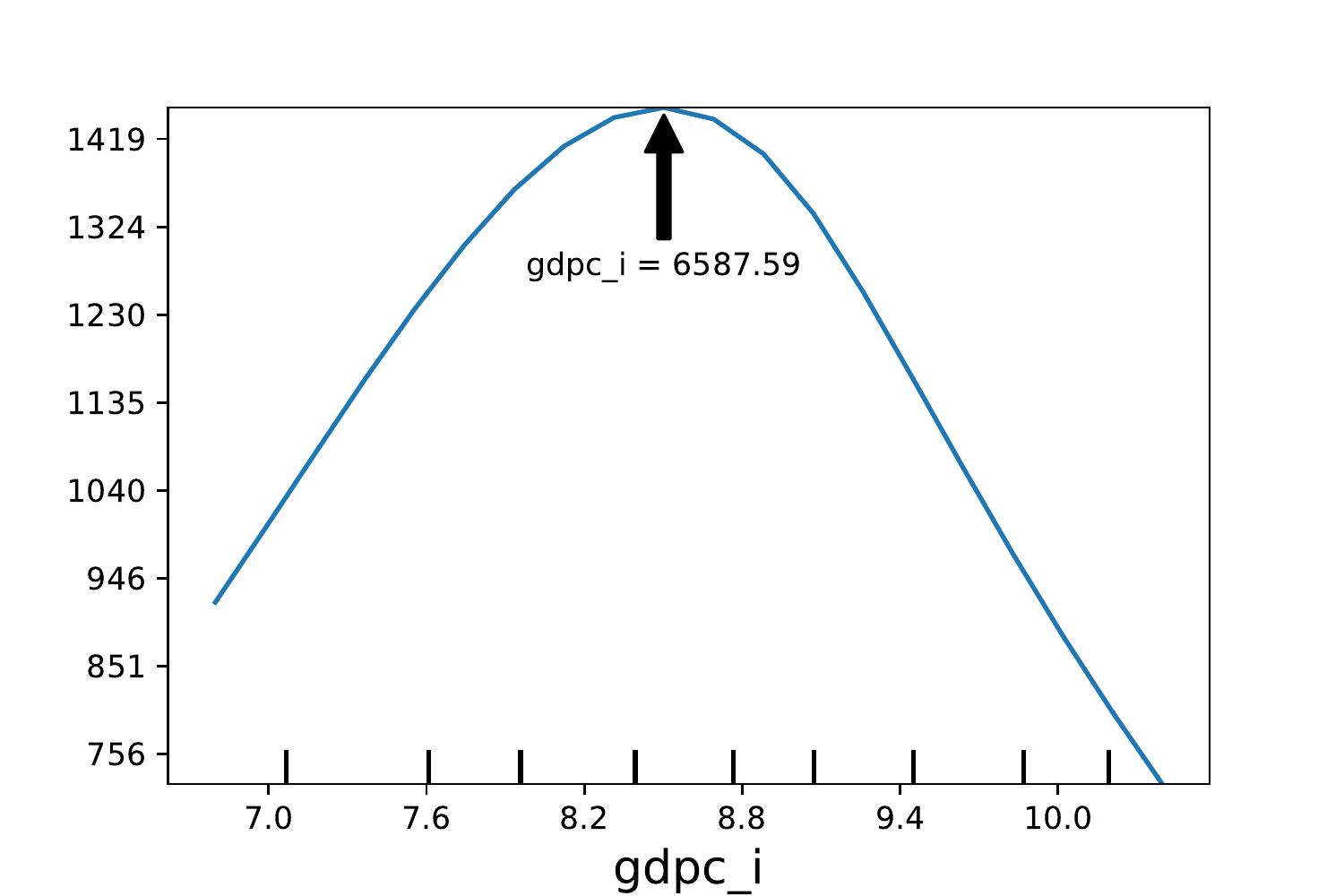}
        \caption{PDP for $gdpc_{i,t}$.}
        \label{fig:pdp_gdpco}
    \end{subfigure}
    ~
    \begin{subfigure}[b]{0.30\textwidth}
        \centering \includegraphics[width=\textwidth]{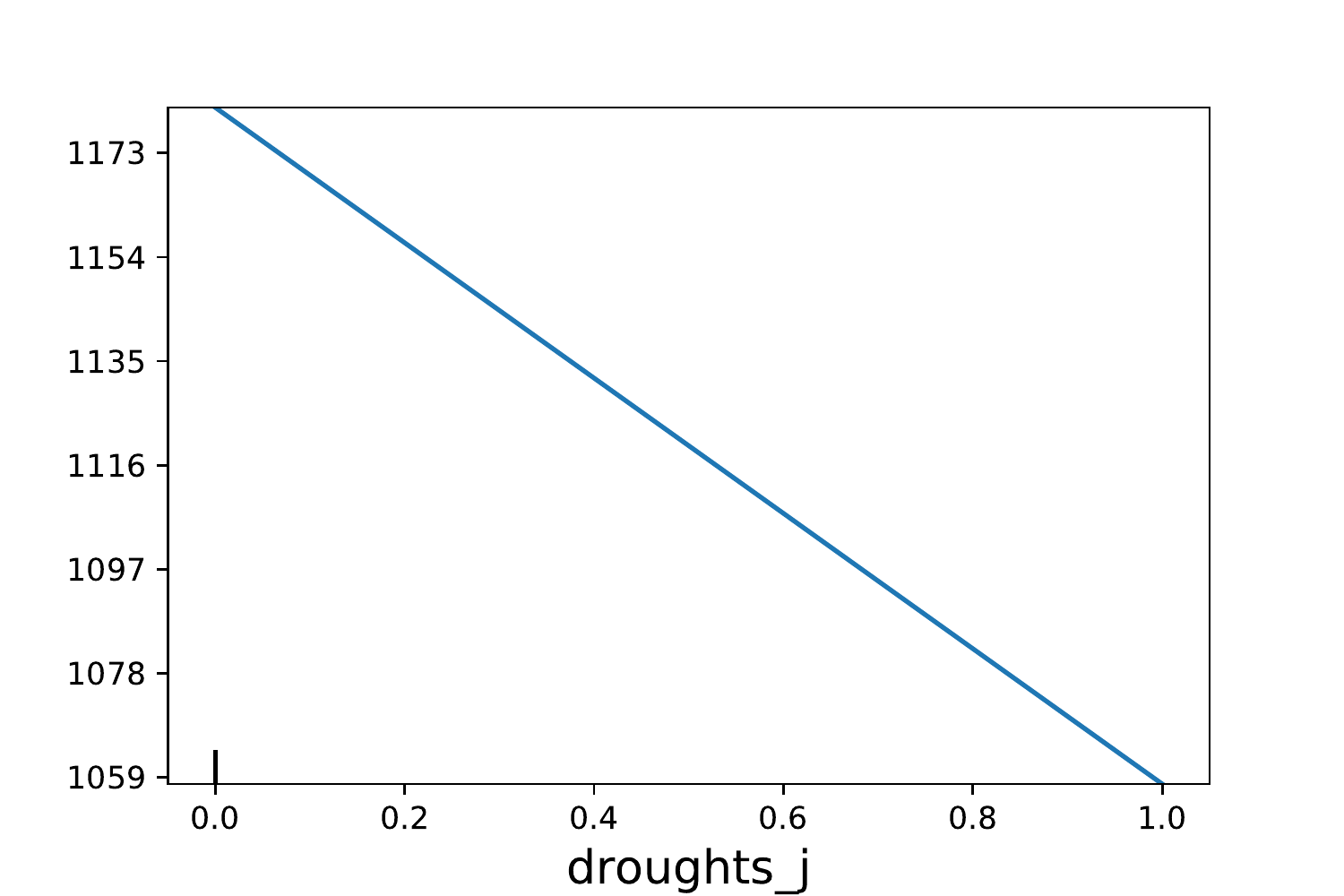}
        \caption{PDP for $droughts_{j,t}$.}
        \label{fig:pdp_droughts}
    \end{subfigure}
    ~
    \begin{subfigure}[b]{0.30\textwidth}
        \centering \includegraphics[width=\textwidth]{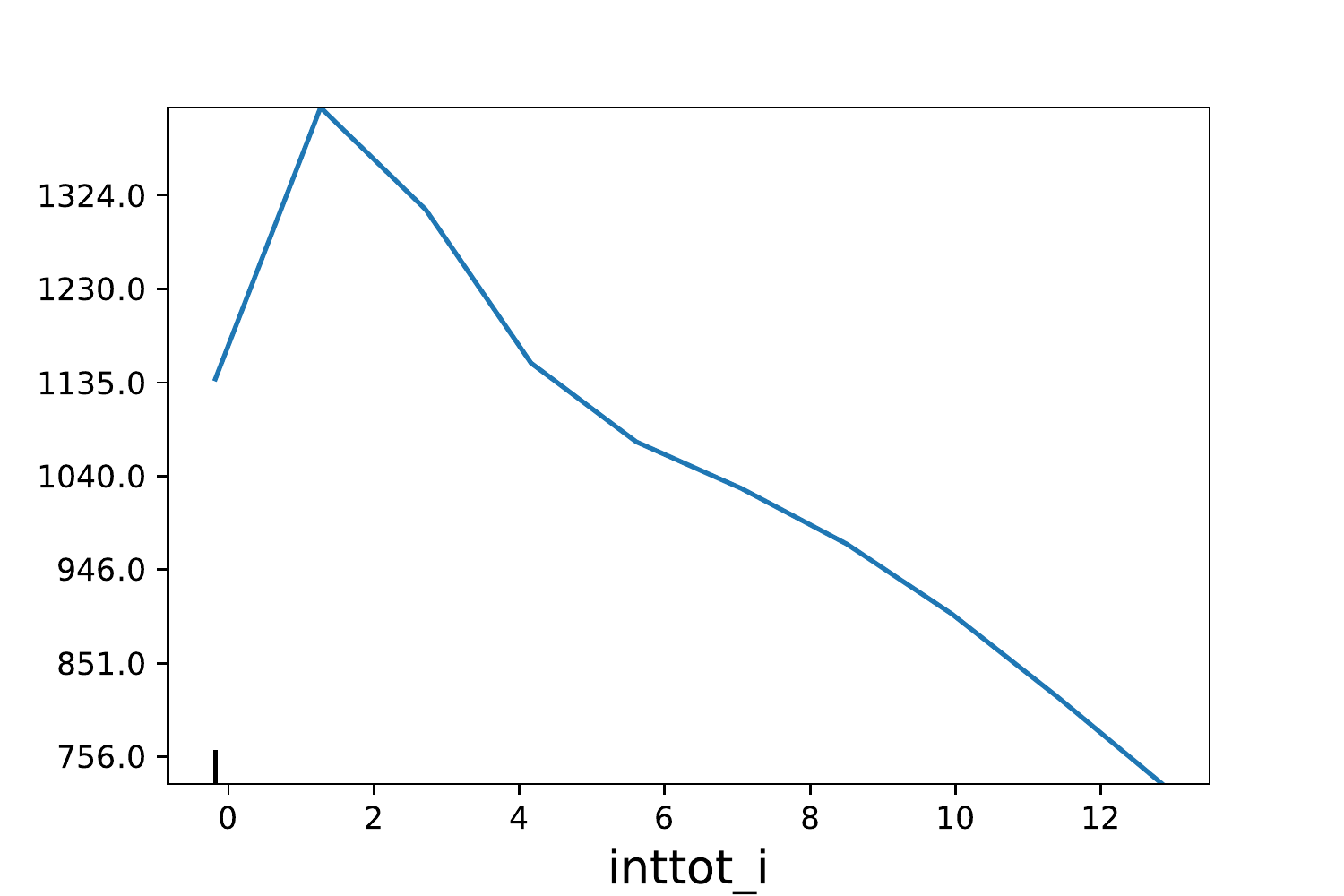}
        \caption{PDP for $inttot_{i,t}$.}
        \label{fig:pdp_inttot}
    \end{subfigure}
    \caption{Features importance and Partial dependence plots for input features. On the top left we have the features importance plot from the Random Forest algorithm. The features values on the PDP are expressed using there natural logarithm values.}

    \label{fig:pdp}
    
\end{figure*}

We use an ANN to estimate the annual migration flows. 
Our ANN is composed of 3 densely connected layers with rectified linear unit (ReLu) activation layers,
which allows us to catch complex, non-linear interactions between the features and flows. 
The output layer uses the sigmoid activation function for the estimation. We use the same model for all predictions $\hat{T}_{i,j}$.
The ANN receives then at each time step the set of features as described in the Table~\ref{tab:features} and returns the forecast migration flow $\hat{T}_{i,j,t+1}$. 

We evaluate the predictive power of the model with several commonly used metrics: the Mean Absolute Error ($MAE$), the Root Mean Square Error ($RMSE$), the Coefficient of determination ($r^2$) and the Common Part of Commuters ($CPC$) also used in \cite{robinsonMachineLearningApproach2017}.
    
\paragraph{Learning process.} To train our model we use three sets. A training set, a validation set, and a test set \cite{goodfellowDeepLearning2016}. 
The training and validation sets consist of the features and flows from $1960$ to $2000$ and the test set consists of the features and flows for the period from $2001$ to $2010$. 
We perform regularization in the model using dropout layers to prevent overfitting and ensure a better generalization \cite{srivastavaDropoutSimpleWay}. 
From the experiments optimal values for the hyperparameters are: number of hidden layer - $3$, number of epoch -  $200$, dropout - $0.10$, batch size - $32$, loss function - $CPC$, found using the $RMSProp$ optimizer \cite{tielemanRmspropDivideGradient2012}.

\paragraph{Partial Dependence Plots.} The partial dependence plots (PDP) \cite{hastieElementsStatisticalLearning2009,molnarInterpretableMachineLearning2018,zhaoCausalInterpretationsBlackBox2019} attempt to better understand the nature of dependence of the approximation $f(features)$ on their joint values. The graphical renderings of the partial dependence provide then the marginal effect of one or two features on the target outcome of a machine learning model.
Consider the  $X_S$ of $l < p$ of the input predictor variables $features =(X_1, X_2, \ldots , X_p)$, indexed by $S \subset \{1, 2, \ldots , p\}$. Let $C$ be the complement set, with $S \cup C = \{1, 2, \ldots , p\}$. Partial dependence functions for the subset $S$ can be estimated by the following equation:
\begin{equation}
\label{eq:pdp}
\centering
    \hat{f}_{S}(x_S)=\frac{1}{n}\sum_{i=1}^n\hat{f}(x_S,x^{(i)}_{C}),
\end{equation}
where $x_S$ are the values of $X_S$ in the training set,  $\{x^{(1)}_{C}, x^{(2)}_{C}, . . . , x^{(n)}_{C}\}$ are the values of $X_C$ occurring in the training data and $n$ is the training data size. 

\section{Experimental Results}
\label{sec:experiments}
The first step of our experiments aims to guarantee the effectiveness of our ML model.
To do so, we compare it to a traditional gravity model estimated by a Poisson Pseudo Maximum-Likelihood approach (PPML)  \cite{beine2016practitioners}. 
As mentioned above, the test set contains all the migration flows from $2001$ to $2010$ representing about $31\%$ of our full sample of data. 
Table~\ref{tab:metrics_cmp} shows the results of the comparison. One can see that the ANN model performs better than the gravity model on all metrics for both the training and  the test sets.
Using the same information, our ANN model is thus more reliable on the task of predicting international migrant flows. 

Next, we analyze how the different features influence the flows. We wish to focus our attention on the most important features. While ideally we would use the ANN to determine which features are important, doing so is challenging, due to the black box nature of ANNs. We decided to use a practical approach in this work in which Random Forests (RFs) are used to filter the initial set of features before analysing these features using more complex approaches. While the performance of this RF is not as good as that of the ANN, this allows to derive some measure of feature importance more easily~ \cite{breimanRandomForests2001}.
Figure~\ref{fig:features_imp} shows the most important features extracted using a RF algorithm. 
As expected, we find that initial migrant stocks are the most important predictors of subsequent migration flows.
Other important features are $distw_{i,j,t}$,  $gdpc_{j,t}$,  $density_{j,t}$,  $pop_{i,t}$,  $pop_{j,t}$, etc.

To understand the actual impact of single features in the ANN, we subsequently use PDPs; these allow for a better understanding of \textit{how} features and outcome are related in the ANN itself. The figures from \ref{fig:pdp_stock} to \ref{fig:pdp_inttot} present the PDP of selected features\footnote{The full set of results is available upon request from the authors.}.
On each figure we represent estimated migration flows $\hat{T}_{i,j,t}$ as a function of the feature we examined. The x-axis represents the natural logarithm of the features $Stock_{i,j,t}$, $distw_{i,j,t}$, $gdpc_{i,t}$ and the absolute value for $droughts_{i,t}$ and $inttot_{i,t}$. 
The y-axis is the flow value estimated by the partial dependence.

The PDP in Figure~\ref{fig:pdp_stock} explains how the initial migrant stocks relate to subsequent flows. International migration flows grow exponentially with the evolution of the migration stock in the destination country. This highlights the potential of migrant networks to foster migration, either through family reunification programs or through easier integration and enhanced flow of information \cite{beine2009diasporas}. 

The second most important feature is $distw_{i,j}$, the population weighted distance between the two countries $i$ and $j$. 
The associated PDP in Figure~\ref{fig:pdp_distw} shows that the farther away two countries are from each other, the lower the respective migrant flow. This reflects the impact of increasing migration costs. 

Next we look at the impact of droughts in destination countries on the flow of incoming migrants. $droughts_{j,t}$ reflects the number of drought month during year $t$ at destination country $j$\footnote{We use the SPEI drought indicator and consider a month as drought month when country aggregated SPEI is at least $1.5$ standard deviations lower than its historic mean \cite{harari2018conflict}.}. Figure~\ref{fig:pdp_droughts} reveals that destination countries tend to receive less migrants during drought years, assumably driven by worse employment possibilities especially for low skilled labour. This example emphasizes the added value of PDPs: while the overall feature importance is low, suggesting to dismiss droughts as migration drivers, the PDP indicates a significant effect in those countries that are concerned.

Figure \ref{fig:pdp_gdpco} reveals a bell shaped relation between GDP per capita at origin and out-going migration flows. This finding is in line with the idea of a mobility transition across different development stages: while the majority of people in poor countries wants to migrate but faces binding financial constraints, the migration incentive decreases once income disparities between origin and destination countries vanish \cite{zelinsky1971hypothesis,dao2018migration}. We find this turning point at a value of 6587 \$ in 2011 chained PPP, roughly reflecting the GDP of Kazakhstan in 2001. This illustrates the potential of PDPs to analyse complex, non-linear relationships between covariates and outcome variables.

In the same vein, Figure \ref{fig:pdp_inttot} shows that the impact of conflicts depends on the conflict intensity. $inttot_{i,t}$ reflects the level of intensity of international conflict during year $t$ for each origin country $i$. While low levels of conflicts push people towards emigration, less and less people migrate when conflicts intensify.

\section{Conclusion}
\label{sec:conclusion}
In this paper, we have proposed an artificial neural network (ANN) model that is able to estimate annual flows better than a classic gravity model. 
With this trained model, we have applied an interpretability technique (partial dependence plots) which allows us to get deeper insights about how migration is influenced by its drivers. PDPs can reveal interesting, non-linear relationship between covariates and outcome variables.
Hence machine learning is not only able to cope with the complex dynamics behind international migration flows, but can also be informative on \textit{how} the flows change w.r.t. the covariates. 
As further work, we believe that it is possible to increase the interpretability by applying other techniques such as Shapley values or LIME with recurrent neural networks.


\bibliographystyle{named}
\bibliography{ijcai20}

\end{document}